# High-Entropy Monoborides: Towards Superhard Materials

Mingde Qin, Qizhang Yan, Haoren Wang, Chongze Hu, Kenneth S. Vecchio, Jian Luo[*]

*Department of NanoEngineering; Program of Materials Science and Engineering, University of California, San Diego, La Jolla, CA, 92093, USA*

**Abstract:**

Single-phase high-entropy monoborides (HEMBs) of the CrB prototype structure have been synthesized for the first time. Reactive spark plasma sintering of ball milled mixtures of elemental precursor powders produced bulk $(V_{0.2}Cr_{0.2}Nb_{0.2}Mo_{0.2}Ta_{0.2})B$, $(V_{0.2}Cr_{0.2}Nb_{0.2}Mo_{0.2}W_{0.2})B$, and $(V_{0.2}Cr_{0.2}Nb_{0.2}Ta_{0.2}W_{0.2})B$ HEMB specimens of ~98.3-99.5% relative densities. Vickers hardness was measured to be ~22-26 GPa at an indentation load of 9.8 N and ~32-37 GPa at 0.98 N. In particular, the load-dependent hardness of $(V_{0.2}Cr_{0.2}Nb_{0.2}Ta_{0.2}W_{0.2})B$ is higher than those of ternary $(Ta_{0.5}W_{0.5})B$ (already considered as superhard) and hardest reported high-entropy metal diborides, and on a par with the classical superhard boride $WB_4$.

**Keywords:** high-entropy ceramics; high-entropy monoborides; reactive sintering; Vickers hardness; superhard materials

---

[*] Corresponding author. E-mail address: jluo@alum.mit.edu (J. Luo).

# Graphical Abstract

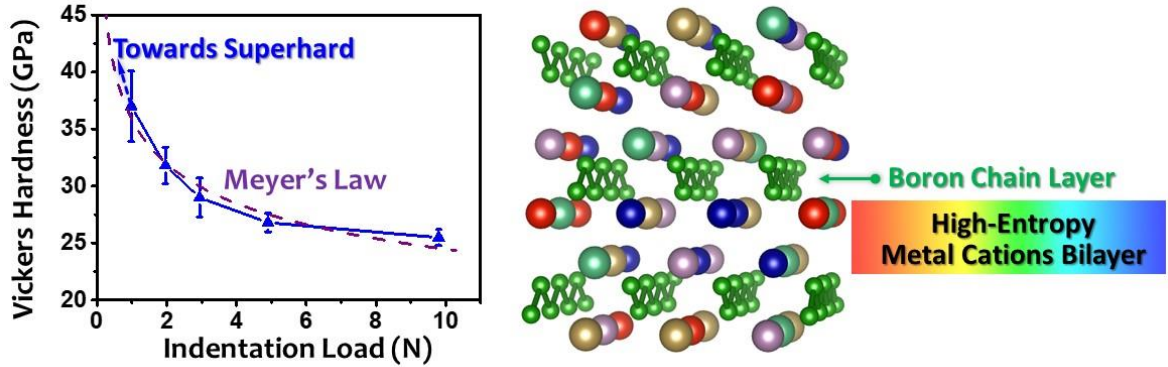

# Highlights:

- First synthesis of high-entropy monoborides (HEMBs)
- Three ~98.3-99.5% dense, single-phase HEMBs were made by reactive SPS
- Vickers hardness values are ~22-26 GPa with an indentation load of 9.8 N
- Vickers hardness values are ~32-37 GPa with an indentation load of 0.98 N
- $(V_{0.2}Cr_{0.2}Nb_{0.2}Ta_{0.2}W_{0.2})B$ is harder than (Ta, W)B and high-entropy metal diborides
- $(V_{0.2}Cr_{0.2}Nb_{0.2}Ta_{0.2}W_{0.2})B$ shows comparable hardness with the superhard $WB_4$
2

The research of high-entropy alloys (HEAs) has received great attentions since the seminal reports by Yeh *et al.* [1] and Cantor *et al.* [2] in 2004. HEAs are widely reported to possess excellent, and sometimes unexpected, mechanical and physical properties [3]. More recently, high-entropy ceramics (HECs), including oxides [4-7], borides [8], carbides [9-11], silicides [12, 13], and fluorides [14] haven been successfully fabricated. Furthermore, the generalized class of compositionally-complex ceramics (CCCs) that also include medium-entropy and/or non-equimolar compositions [15, 16], have been synthesized and studied as the ceramic counterparts to the metallic HEAs and compositionally-complex alloys (CCAs).

Gild *et al.* [8] first reported the successful synthesis of single-phase high-entropy borides (metal diborides) in 2016, as a new class in ultrahigh temperature ceramics (UHTCs). Subsequently, Tallarita *et al.* [17, 18] reported the fabrication of HEBs from elemental precursors via a two-step processing method. Most recently, Qin *et al.* [19] used an in-situ reactive spark plasma sintering (SPS) method, with an isothermal holding prior to final densification to allow out-gassing, to synthesize highly dense HEBs; moreover, they showed that adding softer $MoB_2$ and/or $WB_2$ components to make the high-entropy borides harder. While high-entropy metal diborides have been extensively studied [8, 17, 18, 20-27], the only other high-entropy boride reported to date is a rare earth hexaboride $(Y_{0.2}Yb_{0.2}Sm_{0.2}Nd_{0.2}Eu_{0.2})B_6$ [28].

Specifically, the synthesis of high-entropy monoborides (HEMBs) of the orthorhombic CrB-prototype structure (space group: Cmcm, No. 63) has not been reported. Monoborides are of interest as hard materials, among which ternary $(Ta_{0.5}W_{0.5})B$ has been shown to be "superhard" (*i.e.*, with Vickers hardness ≥ 40 GPa tested at an indentation load of 0.49 N) in a prior study [29]. Only limited materials, including diamond, cubic boron nitride (c-BN), rhenium diboride ($ReB_2$), and tungsten tetraboride ($WB_4$) as the most classic examples, are known to be superhard [30-32]. For borides, the propensity for boron to catenate usually results in extended 2D or 3D network of covalent bonds to present an excellent platform for designing superhard materials [33, 34]. Over the years, superhard materials have been successfully designed and fabricated in monoboride [29, 35], diboride [36], tetraboride [37-39], and dodecaboride [40-43] solid solutions. Notably, CrB-structured ternary $(Ta_xW_{1-x})B$ [29] solid solutions have been found to be superhard at the equimolar composition ($x = 0.5$). This motivated us to further explore HEMBs.

In this study, we have successfully fabricated single-phase CrB-structured HEMBs in bulk form, for the first time to our knowledge, via reactive SPS of elemental metals and boron. The sintered HEMBs achieved ~98.3-99.5% relative densities with virtually no detectable oxides. Specifically, we showed that $(V_{0.2}Cr_{0.2}Nb_{0.2}Ta_{0.2}W_{0.2})B$ is harder than the superhard ternary $(Ta_{0.5}W_{0.5})B$, almost on a par with the classical superhard boride $WB_4$.

To synthesize HEMB, elemental powders of V, Cr, Nb, Mo, Ta, and W (>99.5% purity, ~325 mesh, purchased from Alfa Aesar, MA, USA) and boron (99% purity, 1-2 μm, purchased from US Research Nanomaterials, TX, USA) were utilized for making specimens of three compositions listed in Table 1 as HEMB1 to HEMB3. For each composition, appropriate amounts of metals and boron powders were weighted out in batches of 5 g; 3 at. % of excess boron was added (*i.e.*, a metal-to-boron ratio of 1:1.03) to offset the boron loss due to reaction with native oxide and subsequent evaporation. The powders were first hand-mixed and consecutively high energy ball milled (HEBM) in a Spex 8000D mill (SpexCertPrep, NJ, USA) in tungsten carbide lined stainless steel jars and 11.2 mm tungsten carbide milling media (with a ball-to-powder ratio ≈ 4.5:1) for 50 min with 1 wt. % (0.05 g) of stearic acid as lubricant. The HEBM was performed in an argon atmosphere ($O_2$ < 10 ppm) to prevent oxidation. The milled powders were loaded into 10 mm graphite dies lined with graphite foils in batches of 3 g, and consecutively consolidated into dense pellets via SPS in vacuum ($10^{-2}$ Torr) using a Thermal Technologies 3000 series SPS (CA, USA). The detailed reactive SPS sintering procedure was specified in our prior report [19]. It is noted that specimens were held at 1400°C and then 1600°C, isothermally, before final densification at 2000°C.

All sintered specimens were ground to remove the carbon-rich surface layer from the graphite tooling and polished for characterization. Densities were measured by Archimedes' method, and theoretical densities were calculated from the ideal stoichiometry and the lattice parameters measured by X-ray diffraction (XRD) listed in Supplementary Table S3(a). XRD characterizations were conducted on a Rigaku Miniflex diffractometer with Cu Kα radiation at 30 kV and 15 mA. Scanning electron microscopy (SEM), energy dispersive X-ray spectroscopy (EDS), and electron backscatter diffraction (EBSD) data were obtained with a Thermo-Fisher (formerly FEI) Apreo microscope equipped with an Oxford N-Max[N] EDX detector and an Oxford Symmetry EBSD detector. Vickers microhardness tests were carried out on a LECO diamond microindentor with loading force from 0.98 N (100 gf) to 9.8 N (1 kgf) with holding time of 15 seconds, abiding by ASTM Standard C1327. Over 20 measurements at different locations were conducted for each specimen at each indentation load to ensure the statistical validity and minimize the microstructural and grain boundary effects; specifically, over 50 measurements were conducted for each specimen at 9.8 N indentation load for accuracy. The twin boundary structure and nanoscale elemental distributions were characterized using aberration-corrected scanning transmission electron microscopy (AC-STEM) and EDS using a JEOL 300CF microscope operated at 300 kV. The TEM lamella was prepared with FEI Scios focused ion beam/scanning electron microscope (FIB/SEM).

XRD shows that all three specimens (HEMB1 to HEMB3) synthesized via reactive SPS demonstrate virtually single orthorhombic HEMB phases of the CrB-prototype structure without detectable other boride or oxide phase; however, a tiny amount of rocksalt carbide is detected, which presumably formed due to contamination from SPS graphite tooling (Fig. 1(b)). XRD patterns of the as-milled powders (Fig. 1(a)) show multiple distinct cubic phases, thereby indicating the formation of HEMB phases during the reactive SPS. Peaks broadening observed in Fig. 1(a) can be ascribed to particle grain size reduction, as well as micro-strains caused by HEBM.

For VB and VIB refractory metals, V, Cr, Nb, and Ta form CrB-structured monoborides, whereas Mo and W form CrB-structured orthorhombic (referred to as "orthorhombic" for brevity hereafter) monoborides at elevated temperatures but MoB-structured tetragonal monoborides (*i.e.* α-MoB and α-WB, space group: I41/amd, No.141) at ambient temperature [34, 44, 45]. Both structures are composed of stacking metal bilayers and boron chain layers. Although the orthorhombic β-MoB and β-WB are not thermodynamically stable at ambient temperature, HEMBs with 20 or 40 mol. % MoB and/or WB form the orthorhombic phases. This observation is not a surprise since a prior study found that even 1 cat. % of Ta in $(Ta_{0.01}W_{0.99})B$ can stabilize the solid solution to the orthorhombic phase [29].

The homogeneous distributions of all metal elements in HEMB3 are confirmed by STEM-EDS elemental maps at the nanoscale (Fig. 2(a)) as well as SEM-EDS elemental maps at microscale (Fig. 2(c)). SEM-EDS maps showing homogenous elemental distributions in HEMB1 and HEMB2 are documented in Supplementary Fig. S1. Additional SEM images at lower magnifications are shown in Supplementary Fig. S2. These SEM results further confirm that the specimens are dense with <2 vol. % porosity, consistent with the measured relative densities of >98% for all specimens (Table 1). Averaged lattice parameters calculated by rule-of-mixture (RoM) from individual binary monoborides are also given in the Supplementary Table S3(a), which agree those measured by XRD (with <1% differences). The measured metal compositions by quantitative EDS analysis are $(V_{0.19}Cr_{0.21}Nb_{0.21}Mo_{0.21}Ta_{0.18})B$ for HEMB1, $(V_{0.19}Cr_{0.20}Nb_{0.20}Mo_{0.22}W_{0.19})B$ for HEMB2, and $(V_{0.19}Cr_{0.19}Nb_{0.22}Ta_{0.21}W_{0.19})B$ for HEMB3, where the differences (≤1-2 cat. %) from the equimolar compositions are comparable with EDS measurement errors. Hence, the nominal equimolar compositions are used for further discussion.

Thus, the combination of above results shows that reactive SPS of elemental metals and boron lead to the formation of three orthorhombic HEMBs (listed in Table 1). These represent a new high-entropy phase that was made for the first time (and the third high-entropy boride phases reported [8, 28]).

The reactive SPS procedure ensures the fabrication of dense single-boride-phase specimens, which also represents a novel contribution. The reaction between metals and boron (to form HEMB) is exothermal, which likely takes place during the initial temperature ramping, similar to prior synthesis of metal diborides [46, 47]. Here, the introduction of isothermal holding at 1400°C and 1600°C before final densification, promotes chemical reduction and outgassing of native oxides with the addition of excess boron. On the one hand, the addition of excess boron maintains a local reducing environment during both HEBM and SPS [17]; on the other hand, VB and VIB transition metals utilized can readily form various boron compounds (*e.g.* $M_3B_2$, $M_5B_6$, $M_3B_4$, and $M_2B_3$, where *M* represents a metal [34]) besides monoboride. Therefore, the excess amount of boron has been optimized to be 3% in this study to avoid/minimize the presence/formation of both oxides and other B-rich borides; this also helps achieve high relative densities (~98.3-99.5%). In such case, this optimized process also represents a meaningful technical contribution.

EBSD was utilized to measure the grain size and examine the texture of all sintered specimens. Conducted on well-polished specimen surface normal to SPS pressing direction, EBSD inverse pole figure orientation maps and their corresponding grain size distributions are shown in Supplementary Fig. S3. The averaged grain sizes are measured to be 14.7 ± 9.0 μm for HEMB1, 18.9 ± 14.6 μm for HEMB2, and 13.4 ± 7.9 μm for HEMB3. The average grain size of HEMB2, with a significant amount of grains larger than 40 μm, is appreciably larger than those of HEMB1 or HEMB3.

Notably, twin boundaries (with a misorientation angle of ~41.5 °, measured >30 distinct twin boundaries via EBSD, as well as directly from a STEM micrograph shown in Supplementary Fig. S6) are pervasive in all HEMBs. Multiple grains repeated alternately on the same twinned plane suggest its nature of polysynthetic twinning [48]. Further study is needed to understand the observed microstructures and their formation mechanism.

All HEMBs demonstrate a texture with a depletion of grain orientation of (001) normal to the SPS pressing direction. This texture is also obvious in the inverse pole figures of the crystal preferred orientation shown in Supplementary Fig. S4, and it can be further confirmed by comparing the measured XRD patterns of sintered pellets (Fig. 1(b)) with the calculated XRD patterns assuming totally random cation occupation and grain orientation (Supplementary Fig. S5). We believe the formation of this texture in our HEMB specimens is related to the applied pressure in SPS, as a similar texture was not observed in prior $(Ta_xW_{1-x})B$ solid solutions synthesized by arc-melting [29].

In addition to the first synthesis of HEMBs, a second significant finding of this study is the enhanced hardness of the HEMBs, in comparison with the ternary $(Ta_{0.5}W_{0.5})B$ that is already considered to be a "superhard" material [29]. Fig. 3 shows the measured Vickers microhardness of HEBMs under different indentation loads from 0.98 N to 9.8 N. At a high indentation load of 9.8 N, hardness is measured to be 24.3 ± 0.6 GPa for HEMB1, 22.3 ± 0.6 GPa for HEMB2, and 25.5 ± 0.8 GPa for HEMB3. The measured hardness increases steadily with decreasing indentation load, reaching 34.1 ± 2.3 GPa for HEMB1, 32.6 ± 2.5 GP for HEMB2, and 37.0 ± 3.1 GPa for HEMB1, at an indentation load of 0.98 N.

This load-dependent hardness (*a.k.a.* indentation size effect [49]) is usually observed in brittle and incompressible materials with high hardness [29, 30], which has been ascribed to mixed elastic/plastic response during plastic deformation [50] and friction (or the underlying surface-to-volume ratio) of indentation [51].

Limited by the optical system on our hardness tester, we could not measure the Vickers hardness at a lower load of 0.49 N for our HEMBs accurately (because indentations are too small to be precisely measured). To estimate the hardness at lower loads, Meyer's law [52-54] was applied to fit the relationship between hardness and indentation: $P = a \cdot d^n$, where *P* is the indentation load (g), *d* is the averaged indentation length (μm), *n* is an index, and *a* is a constant.

The best-fitted curves are represented by the purple dash lines in Fig. 3. The corresponding fitted parameters and regression analyses results are given in Supplementary Table S1. Based on this relationship, the hardness at the 0.49 N indentation load are estimated to be 35.7 GPa for HEMB1, 35.2 GPa for HEMB1, and 40.2 GPa for HEMB3. Thus, HEMB3 ($V_{0.2}Cr_{0.2}Nb_{0.2}Ta_{0.2}W_{0.2}$)B may be considered as a superhard material based on the somewhat artificial criterion of $H_v \geq 40$ GPa that was typically measured at 0.49 N in the prior studies of superhard boride materials [29, 35-43]. We should note that this cut-off criterion is somewhat subjective and perhaps not the most important issue here.

To further benchmark the hardness of HEMB3 ($V_{0.2}Cr_{0.2}Nb_{0.2}Ta_{0.2}W_{0.2}$)B, we compare its load-dependent hardness with those of the two previously reported superhard materials, namely the classical $WB_4$ [37] and the ternary ($Ta_{0.5}W_{0.5}$)B [29] in Supplementary Fig. S7. Under the same indentation loads from 0.98 N to 4.9 N, the measured hardness values of HEMB3 evidently fall between those of $WB_4$ and ($Ta_{0.5}W_{0.5}$)B.

Notably, HEMB3 ($V_{0.2}Cr_{0.2}Nb_{0.2}Ta_{0.2}W_{0.2}$)B is substantially harder than the ternary ($Ta_{0.5}W_{0.5}$)B that was considered as a superhard material [29] in the entire load range of 0.98N to 4.9N, as shown in Supplementary Fig. S7, which is probably more significant practically. To eliminate the possible effects from different fabrication routes, an additional specimen of ($Ta_{0.5}W_{0.5}$)B is fabricated via the same metal-boron reactive SPS method used to fabricate HEMBs. This ($Ta_{0.5}W_{0.5}$)B specimen prepared by SPS in this study (Fig. 3(d)) shows similar Vickers microhardness as the one made by arc melting in Ref. [29]; see Supplementary Fig. S8 for the comparison and additional characterization results of the ($Ta_{0.5}W_{0.5}$)B specimen made in this study. Hence, HEMB3 is harder than the ternary ($Ta_{0.5}W_{0.5}$)B regardless the fabrication routes.

This observation is also scientifically interesting as it suggests that the hardness can be further enhanced in a high-entropy solution in comparison with a (perhaps the best) ternary subsystem. This is even more interesting scientifically because all the constituent binary monoborides (except VB, whose measured hardness is unavailable in literature) have noticeably lower hardness [34, 55] (*viz.* CrB: 19.6 GPa at 9.8 N, NbB: 21.5 GPa at 0.49 N, MoB: 23-25 GPa at 0.49 N, TaB: 30.7 GPa at 0.49 N, and WB: 36.3 GPa at 0.49 N). In other words, we can view this case as adding substantial amounts of three softer components (into ($Ta_{0.5}W_{0.5}$)B to form the high-entropy ($(V_{0.2}Cr_{0.2}Nb_{0.2}Ta_{0.2}W_{0.2})B$) makes the HEMB harder, akin to what we have recently observed in high-entropy metal diborides [19].

Finally, we further compare the measured Vickers hardness of these HEMBs with several (probably the hardest reported) high-entropy metal diborides synthesized via a similar reactive SPS route in our prior study [19] in Supplementary Table S2. At the same indentation load of 1.96 N, the HEMBs made in this study are notably harder than those hardest reported high-entropy metal diborides (27.2-31.8 GPa *vs.* 24.9-27.5 GPa shown in Supplementary Table S2).

In summary, this study demonstrates the first synthesis of HEMBs via a novel metal-boron reactive SPS procedure from elemental boron and metals. We have successfully fabricated HEMBs in bulk form with high relative densities (98.3-99.5%) with virtually no observable oxides inclusions and minimal impurities. The sintered HEMBs exhibit high hardness. Specifically, the most promising composition ($(V_{0.2}Cr_{0.2}Nb_{0.2}Ta_{0.2}W_{0.2})B$) exhibits load-dependent hardness that falls between those of the two superhard materials, namely $WB_4$ [37] and $(Ta_{0.5}W_{0.5})B$ [29], reported previously. It is also scientifically interesting to suggest that the hardness can be further enhanced in a high-entropy solution in comparison with their ternary subsystems. It is interesting to further note that the density of HEMB3 is only ~2/3 of that of the ternary $(Ta_{0.5}W_{0.5})B$ [29] (Table 1; an advantage for certain applications), yet it is harder than the already superhard $(Ta_{0.5}W_{0.5})B$.

Acknowledgement: This work is supported by an Office of Naval Research MURI program (Grant No. N00014-15-1-2863; Program Managers: Dr. Eric Wuchina and Dr. Kenny Lipkowitz).

**Table 1.** Summary of three HEMB specimens studied, along with a reference ternary $(Ta_{0.5}W_{0.5})B$ specimen made by the same procedure in this study. Theoretical densities were calculated from the measured lattice parameters and compositions. The specimen densities were measured via the Archimedes method. See Supplementary Table S3 for lattice parameters. Averaged grain sizes were obtained from EBSD analyses shown in Supplementary Fig. S3.

| Specimen | Compositions | Theoretical Density (g/cm³) | Measured Density (g/cm³) | Relative Density | Vickers Hardness (GPa) at 9.8 N Load | Vickers Hardness (GPa) at 0.98 N Load | Grain Size (μm) |
|---|---|---|---|---|---|---|---|
| HEMB1 | $(V_{0.2}Cr_{0.2}Nb_{0.2}Mo_{0.2}Ta_{0.2})B$ | 8.60 | 8.55 | 99.4% | 24.3 ± 0.6 | 34.1 ± 2.3 | 14.7 ± 9.0 |
| HEMB2 | $(V_{0.2}Cr_{0.2}Nb_{0.2}Mo_{0.2}W_{0.2})B$ | 8.84 | 8.69 | 98.3% | 22.3 ± 0.6 | 32.6 ± 2.5 | 18.9 ± 14.6 |
| HEMB3 | $(V_{0.2}Cr_{0.2}Nb_{0.2}Ta_{0.2}W_{0.2})B$ | 10.00 | 9.95 | 99.5% | 25.5 ± 0.8 | 37.0 ± 3.1 | 13.4 ± 7.9 |
| Reference | $(Ta_{0.5}W_{0.5})B$ | 14.97 | 14.72 | 98.2% | 21.2 ± 0.8 | 32.9 ± 1.7 | |

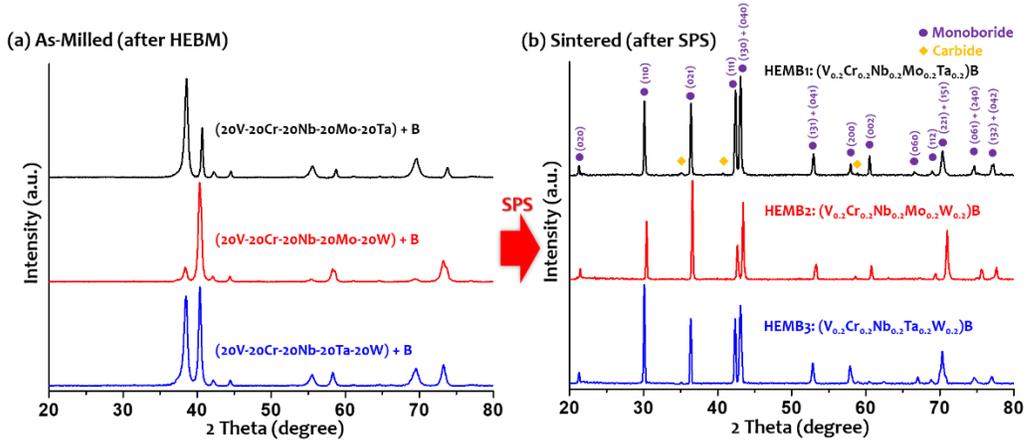

**Fig. 1.** XRD patterns of specimens synthesized and fabricated via HEBM and reactive SPS: **(a)** as-milled metal and boron powders mixtures (after HEBM) and **(b)** sintered HEMB pellets (after SPS). All three sintered HEMB specimens demonstrate a largely single CrB-structured orthorhombic phase, albeit a tiny amount of rocksalt carbide is detected due to contamination from graphite tooling during SPS. Note that some diffraction peaks in CrB-orthorhombic phase are overlapped and the XRD patterns of as-milled powders are not indexed because of extensive peak overlapping of six elements.

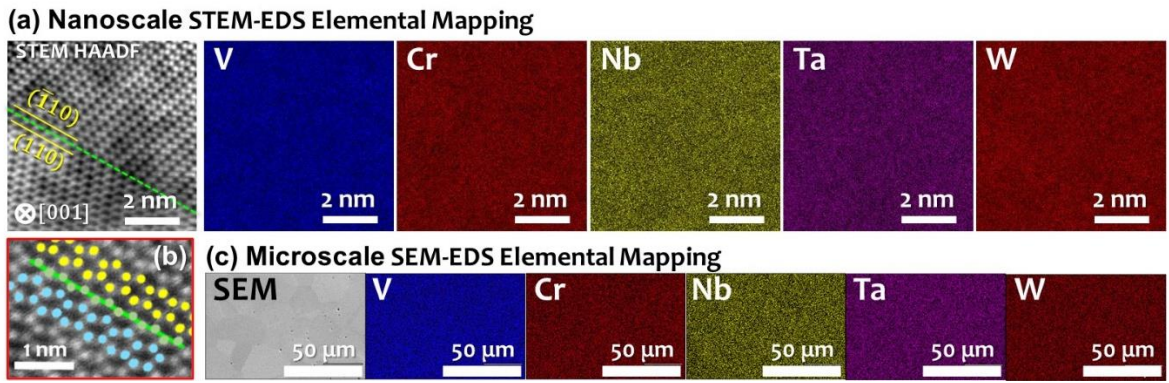

**Fig. 2.** (a) STEM micrograph and corresponding EDS elemental maps of specimen HEMB3 $(V_{0.2}Cr_{0.2}Nb_{0.2}Ta_{0.2}W_{0.2})B$. The STEM HAADF image also shows a $(110)//(\bar{1}10)$ twin boundary (indicated by the green dashed line). (b) An enlarged STEM HAADF image showing the twin boundary where atom positions of the two bordering grains are labeled by blue and yellow dots, respectively. See Supplementary Fig. S6 for additional STEM micrographs of the twin boundary and analysis. (c) SEM micrograph and corresponding EDS elemental maps of the same specimen are also presented. The compositions are homogeneous at both microscale and nanoscale. See Supplementary Fig. S1 for the SEM-EDS elemental maps of specimens HEMB1 and HEMB2.

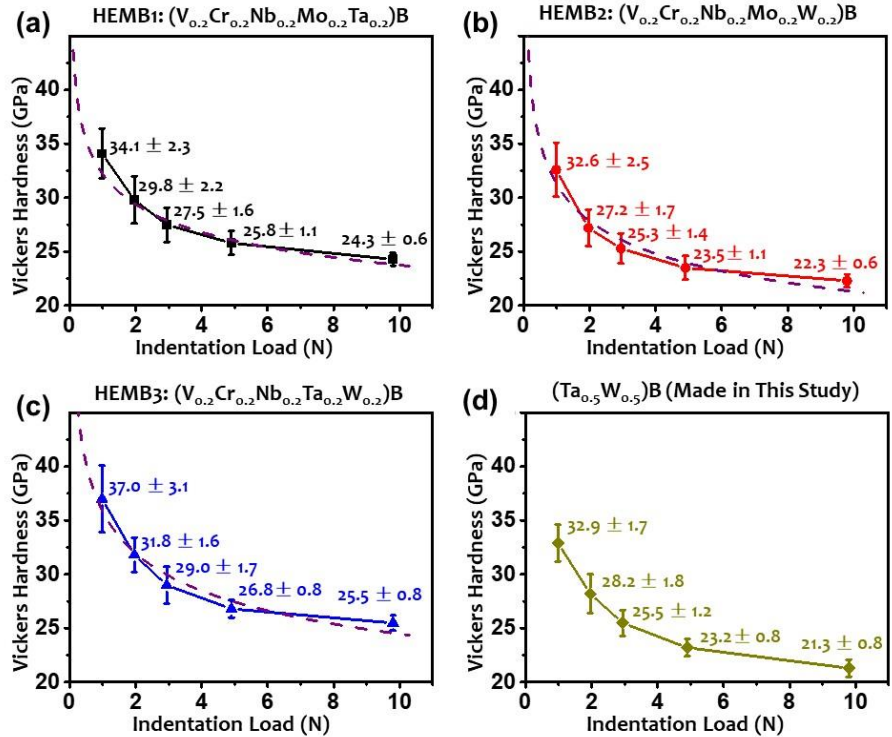

**Fig. 3.** Measured Vickers microhardness of **(a)** HEMB1, **(b)** HEMB2, and **(c)** HEMB3, along with **(d)** a reference $(Ta_{0.5}W_{0.5})W$ specimen made in this study by the same procedure, under indentation loads ranging from 0.98 N to 9.8 N. The purple dash lines represent the best-fitted microhardness vs. indentation load curves by Meyer's law for each specimen. See Supplementary Table S1 for the regression analyses to fit Meyer's law.

# Supplementary Data

# High-Entropy Monoborides: Towards Superhard Materials

Mingde Qin, Qizhang Yan, Haoren Wang, Chongze Hu, Kenneth S. Vecchio, Jian Luo*

Department of NanoEngineering; Program of Materials Science and Engineering
University of California, San Diego
La Jolla, CA, 92093, USA

* Corresponding author. E-mail address: jluo@alum.mit.edu (J. Luo).

**Table S1.** Regression analysis results of the measured hardness *vs.* indentation load curves according to Meyer's law. All parameters are fitted based on Meyer's law [1-3]:

$$P = a \cdot d^n$$

or:
$$\log P = n \cdot \log d + \log a,$$

where $P$ is the indentation load (g), $d$ is the averaged indentation length (μm), $n$ is an index, and $a$ is a constant.

| Specimen | n | log a | Correlation Factor $r^2$ |
|---|---|---|---|
| HEMB1 | 1.7665 | 0.4530 | 0.9994 |
| HEMB2 | 1.7152 | 0.4875 | 0.9987 |
| HEMB3 | 1.7164 | 0.5374 | 0.9990 |

**Table S2.** Comparison of the measured Vickers microhardness of high-entropy monoborides (HEMBs) in this study and the (hardest reported) high-entropy metal diborides synthesized via similar reactive SPS method in a prior study [4]. The Vickers microhardness listed here were all measured at the same indentation load of 1.96 N. Measured densities, as well as the relative densities, are also listed for reference.

| High-entropy Monoboride Compositions | Density (g/cm$^3$) (Relative Density) | Vickers Hardness at 1.96 N (GPa) |
|---|---|---|
| HEMB1: $(V_{0.2}Cr_{0.2}Nb_{0.2}Mo_{0.2}Ta_{0.2})B$ | 8.55 (99.4%) | 29.8 ± 2.2 |
| HEMB2: $(V_{0.2}Cr_{0.2}Nb_{0.2}Mo_{0.2}W_{0.2})B$ | 8.69 (98.3%) | 27.2 ± 1.7 |
| HEMB3: $(V_{0.2}Cr_{0.2}Nb_{0.2}Ta_{0.2}W_{0.2})B$ | 9.95 (99.5%) | 31.8 ± 1.6 |
| **High-entropy Diboride Compositions** | | |
| $(Ti_{0.2}Zr_{0.2}Nb_{0.2}Mo_{0.2}Ta_{0.2})B_2$ | 7.46 (99.2%) | 24.9 ± 1.3 |
| $(Ti_{0.2}Zr_{0.2}Mo_{0.2}Hf_{0.2}W_{0.2})B_2$ | 8.35 (97.5%) | 26.0 ± 1.5 |
| $(Zr_{0.2}Nb_{0.2}Hf_{0.2}Ta_{0.2}W_{0.2})B_2$ | 9.78 (98.1%) | 26.7 ± 1.1 |
| $(Zr_{0.225}Mo_{0.225}Hf_{0.225}Ta_{0.225}W_{0.1})B_2$ | 9.56 (98.8%) | 27.5 ± 1.1 |

**Table S3.** Summary of lattice parameters

**Table S3(a).** Summary of the lattice parameters of HEMBs made in this study. Lattice parameters were measured by XRD, whereas rule-of-mixture (RoM) averages were calculated from data of binary metal monoborides shown in Part (b).

| Specimen | Compositions | Measured Lattice Parameters by XRD $a, b, c$ (Å) | RoM Averaged Lattice Parameters $a, b, c$ (Å) |
|---|---|---|---|
| HEMB1 | $(V_{0.2}Cr_{0.2}Nb_{0.2}Mo_{0.2}Ta_{0.2})B$ | 3.178, 8.375, 3.056 | 3.150, 8.359, 3.059 |
| HEMB2 | $(V_{0.2}Cr_{0.2}Nb_{0.2}Mo_{0.2}W_{0.2})B$ | 3.150, 8.301, 3.045 | 3.122, 8.318, 3.042 |
| HEMB3 | $(V_{0.2}Cr_{0.2}Nb_{0.2}Ta_{0.2}W_{0.2})B$ | 3.180, 8.374, 3.066 | 3.149, 8.351, 3.058 |

**Table S3(b).** Summary of the lattice parameters of binary monoborides with references. Although the α-MoB and α-WB (I41/amd, No.141) are the thermodynamically stable phases at room temperature, the lattice parameters of the CrB-structured β-MoB and β-WB are listed here.

| Monoboride Phase | Lattice Parameters $a, b, c$ (Å) |
|---|---|
| VB | 3.060, 8.048, 2.972 [5] |
| CrB | 2.978, 7.870, 2.935 [6] |
| NbB | 3.297, 8.722, 3.165 [7] |
| β-MoB | 3.142, 8.496, 3.072 [8] |
| TaB | 3.275, 8.660, 3.153 [9] |
| β-WB | 3.135, 8.454, 3.066 [10] |

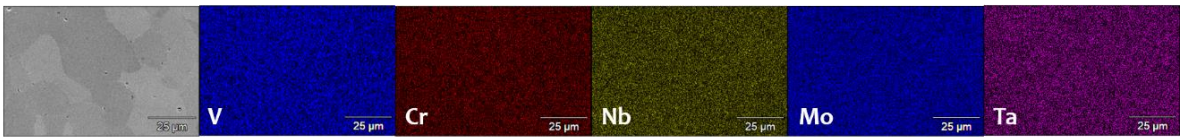
(a) HEMB1: $(V_{0.2}Cr_{0.2}Nb_{0.2}Mo_{0.2}Ta_{0.2})B$

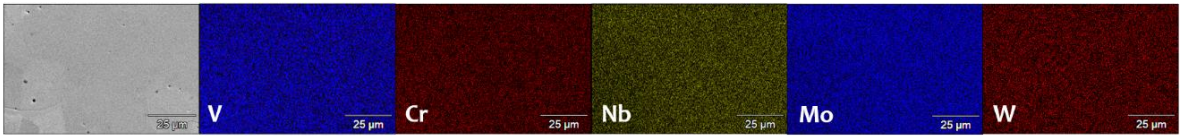
(b) HEMB2: $(V_{0.2}Cr_{0.2}Nb_{0.2}Mo_{0.2}W_{0.2})B$

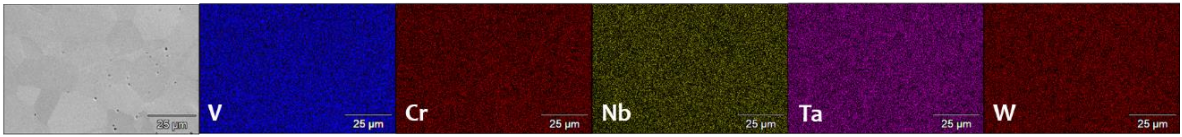
(c) HEMB3: $(V_{0.2}Cr_{0.2}Nb_{0.2}Ta_{0.2}W_{0.2})B$

**Fig. S1.** SEM micrographs and corresponding EDS elemental maps of three HEMB specimens **(a)** HEMB1, **(b)** HEMB2, and **(c)** HEMB3. All three sintered specimens show homogeneous elemental distributions.

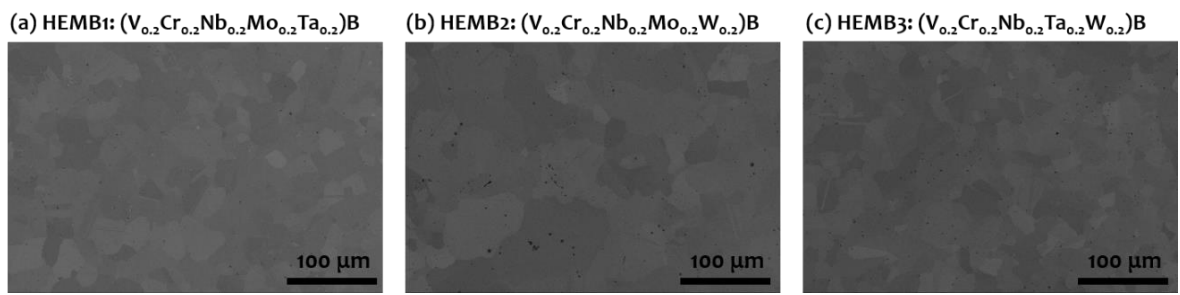

**Fig. S2.** SEM micrographs of three sintered HEMB specimens **(a)** HEMB1, **(b)** HEMB2, and **(c)** HEMB3. The small black spots in the micrographs are from pores.

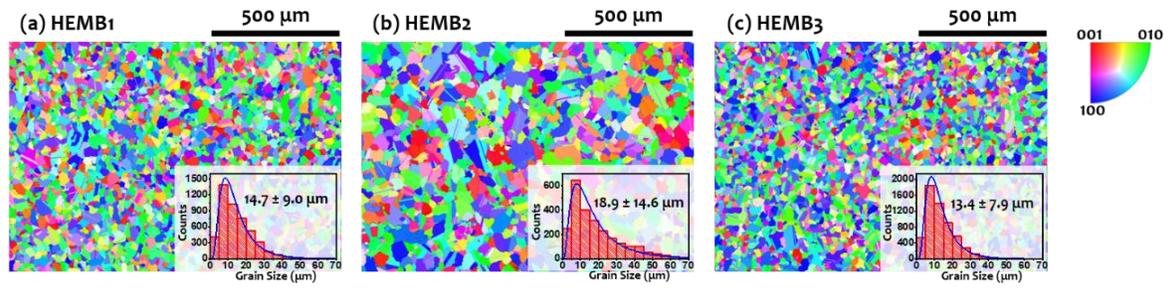

**Fig. S3.** EBSD normal direction inverse pole figure orientation maps for three HEMB specimens on the plane perpendicular to the pressing direction in SPS: **(a)** HEMB1, **(b)** HEMB2, and **(c)** HEMB3. The insets show grain size distributions.

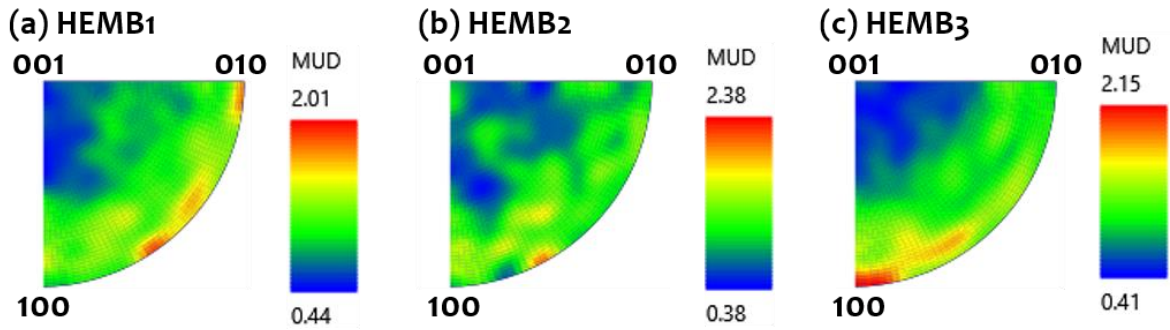

**Fig. S4.** Normal direction inverse pole figures of the crystal preferred orientation measured from EBSD for three sintered specimens: **(a)** HEMB1, **(b)** HEMB2, and **(c)** HEMB3. Contour maps represent multiples of the uniform distribution (MUD). All sintered specimens feature textures with depletion of the (001) plane grain orientation perpendicular to the pressing direction in SPS.

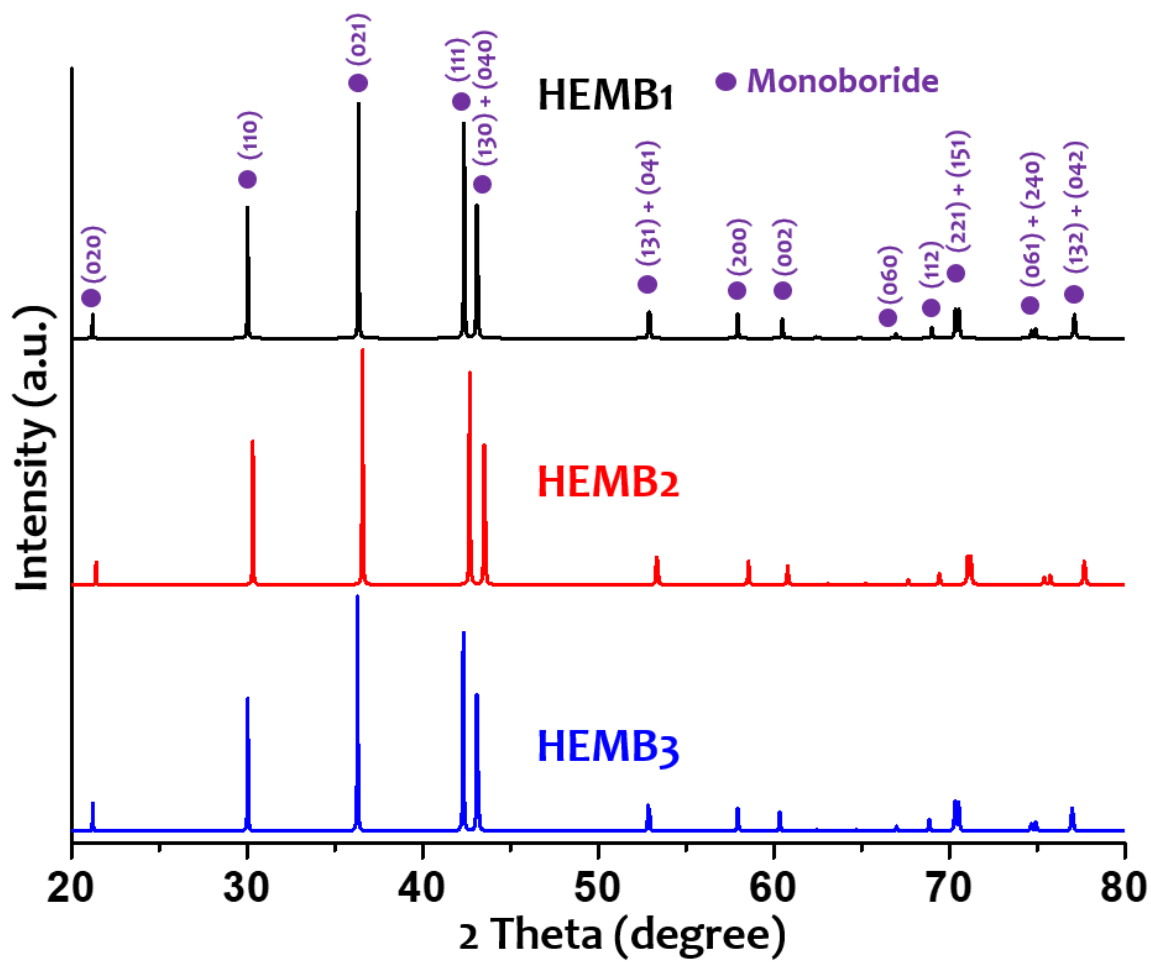

**Fig. S5.** Calculated powder XRD patterns based on the kinematic theory of diffraction for three compositions assuming a random occupation of five cations in the CrB-prototype HEMB crystal structure. Noting that the intensity of (111) peak is slightly lower than that of the maximum (021) peak, and it is significantly higher than that of the (130) + (040) combined peak.

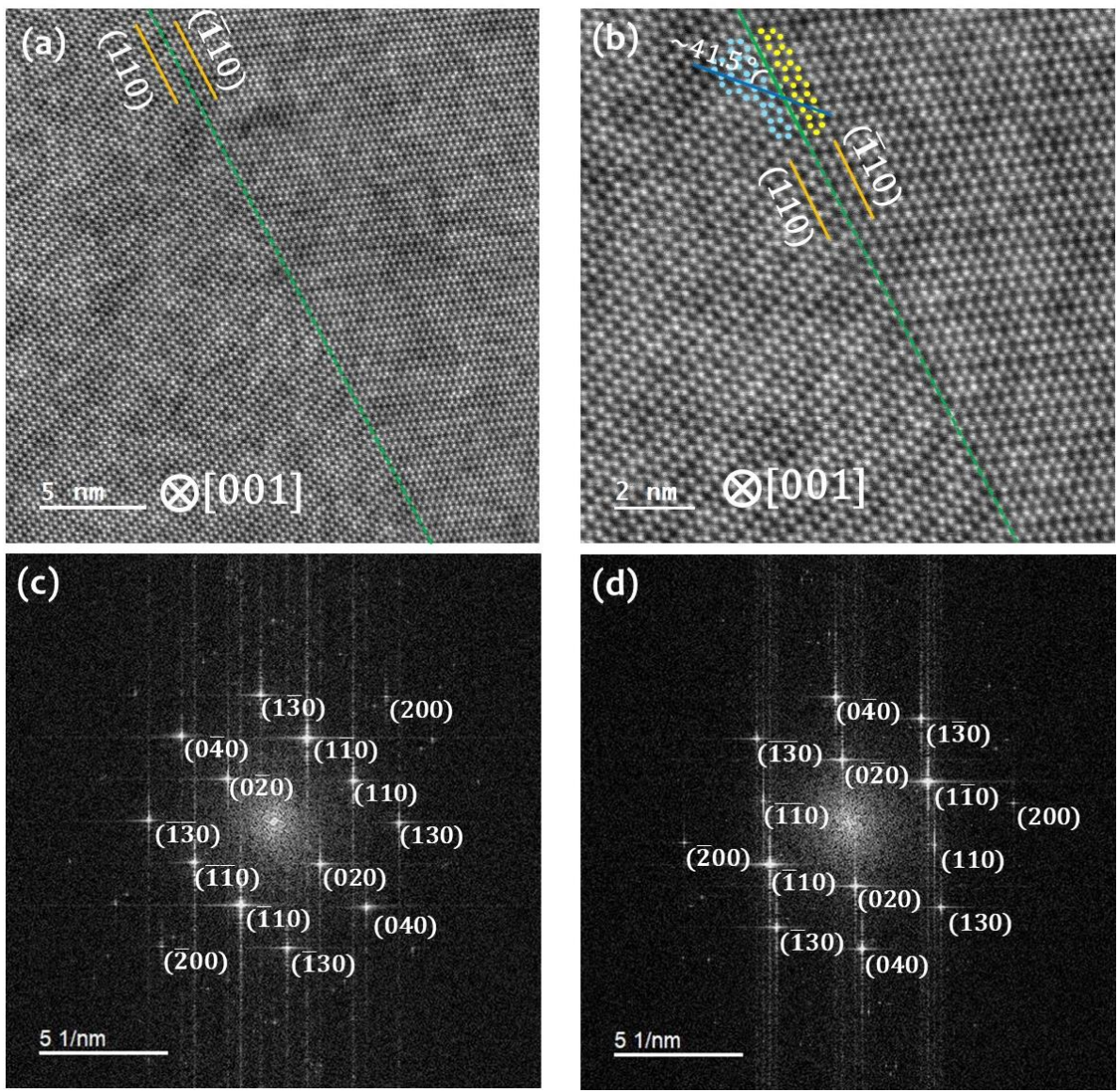

**Fig. S6. (a, b)** STEM micrographs of a $(110)//(\bar{1}10)$ twin boundary on specimen HEMB3 at different magnifications. The green line indicates the twin boundary plane; and the blue line indicates the $(\bar{1}10)$ plane on the left grain. The misorientation angle between two grains is measured to be ~41.5°. **(c, d)** FFT diffraction patterns of the two grains, showing the crystallographic orientations.

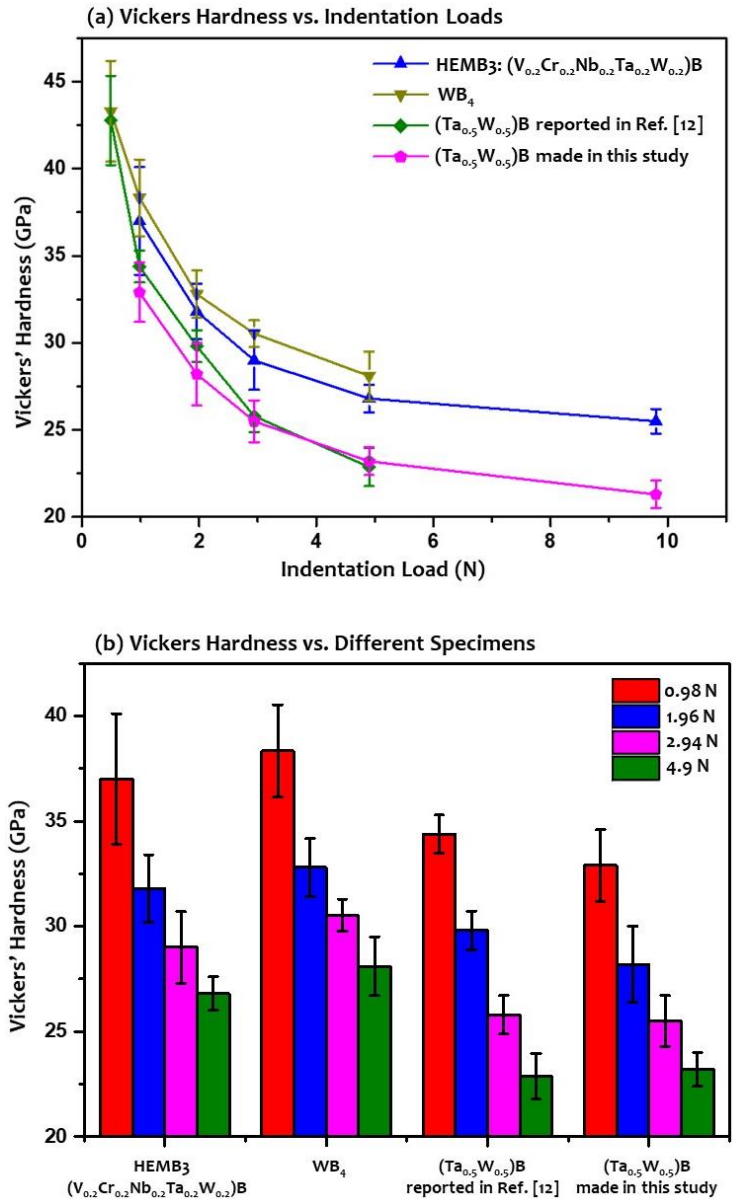

**Fig. S7.** Comparison of measured Vickers microhardness of HEMB3 $(V_{0.2}Cr_{0.2}Nb_{0.2}Ta_{0.2}W_{0.2})B$ and a reference $(Ta_{0.5}W_{0.5})B$ made in this study by identical procedure, with those of two superhard materials, $WB_4$ [11] and $(Ta_{0.5}W_{0.5})B$ [12], reported in literature. **(a)** Vickers hardness *vs.* indentation loads. **(b)** Bar charts of measured Vickers hardness of different specimens at four comparable indentation loads.

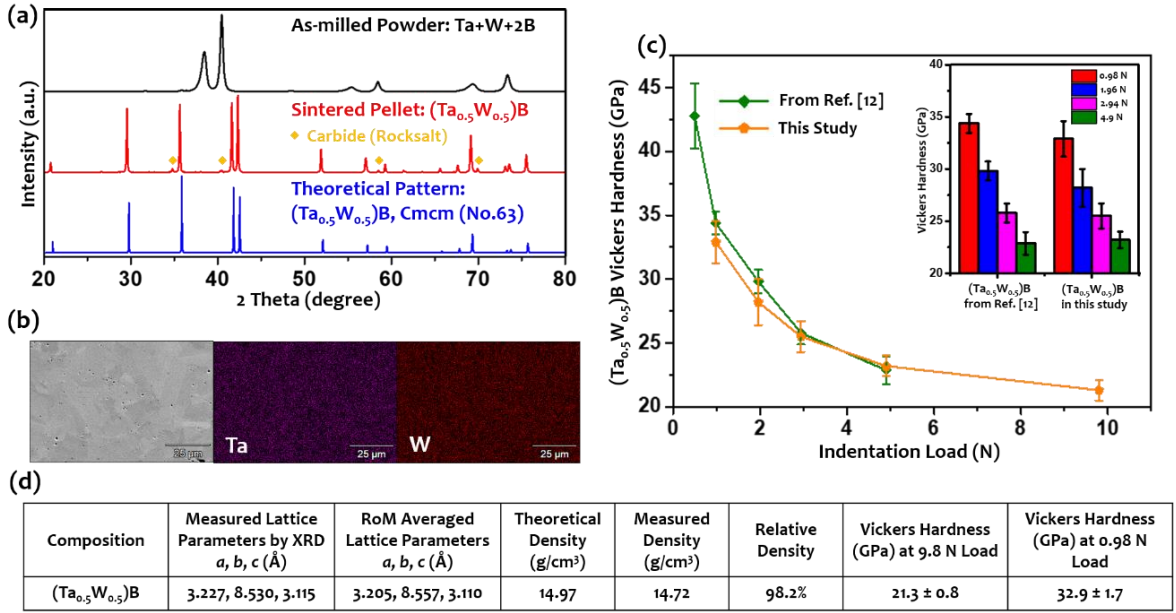

**Fig. S8.** Characterization of $(Ta_{0.5}W_{0.5})B$ specimen synthesized via the identical HEBM and reactive SPS route (as other HEMBs) in this study. **(a)** XRD patterns of $(Ta_{0.5}W_{0.5})B$ specimen as-milled metal and boron powder mixture (black line), sintered $(Ta_{0.5}W_{0.5})B$ pellet (red line), and calculated $(Ta_{0.5}W_{0.5})B$ powder assuming a random occupation of cations in the CrB-prototype structure (blue line). **(b)** SEM micrographs and corresponding EDS elemental maps of $(Ta_{0.5}W_{0.5})B$ specimen. **(c)** Measured Vickers microhardness of $(Ta_{0.5}W_{0.5})B$ specimen made in this study under indentation loads from 0.98 N to 9.8 N (orange line). Vickers microhardness of the reported superhard material with same composition $(Ta_{0.5}W_{0.5})B$ from Ref. [12] under indentation loads from 0.49 N to 4.9 N are also displayed for comparison (green line). The inset is the bar charts of comparing measured Vickers microhardness of $(Ta_{0.5}W_{0.5})B$ made by two different methods. Both $(Ta_{0.5}W_{0.5})B$ specimens possess comparable Vickers microhardness despite different fabrication routes adopted (metal-boron reactive SPS in this study *vs.* arc-melting in Ref. [12]). **(d)** Summary of the properties of $(Ta_{0.5}W_{0.5})B$ specimen synthesized in this study.